\documentclass[amsmath,amssymb]{revtex4}

\usepackage[applemac]{inputenc}
\usepackage{amsfonts}
\usepackage{graphicx}% Include figure files
\usepackage{dcolumn}% Align table columns on decimal point
\usepackage{bm}% bold math
\usepackage[english]{babel}

\newcommand{\lbl}{\label}

\newcommand{\be}{\beqn}
\newcommand{\ee}{\eeqn}
\newcommand{\bea}{\begin{eqnarray}}
\newcommand{\eea}{\end{eqnarray}}
\newcommand{\hf} {{1/2}}

\newcommand{\beqn}{\begin{eqnarray}}
\newcommand{\eeqn}{\end{eqnarray}}
\newcommand{\nn}{\nonumber \\}

\newcommand{\bE}{\mathbf{E}}
\newcommand{\bB}{\mathbf{B}}

\def\Fij#1#2{\mathcal{F}_{#1 #2}}
\def\dFij#1#2{\mathcal{F^*}_{#1 #2}}

\def\eq#1{(\ref{#1})}

\def\hf{\frac{1}{2}}

\def\eq#1{(\ref{#1})}

\def\trm#1{\textrm{#1}}

\def\eq#1{(\ref{#1})}

\def\eq#1{(\ref{#1})}

\def\d#1d#2{\frac{\partial #1}{\partial #2}}

\begin{document}

\preprint{APS/123-QED}

\title{Complex Faraday's Tensor for the Born-Infeld Theory}

\author{Michel Gondran$^a$}\email{michel.gondran@chello.fr}
\author{Abdelouahab Kenoufi$^b$}\email{abdel.kenoufi@unibas.ch}
%\author{Janos Polonyi$^{c,d}$}\email{polonyi@lpt1.u-strasbg.fr}

\affiliation{$^a$Institut National de Recherche en Informatique et Automatique, Rocquencourt, Domaine de Voluceau, B.P. 105, 78153 Le Chesnay, France}

\affiliation{$^b$Institut für Physik, Universität Basel, Klingelbergstrasse 82, 4050 CH-Basel, Switzerland}
%\affiliation{$^c$Laboratoire de Physique Th\'eorique, Universit\'e Louis
%Pasteur\\
%3-5 rue de l'Universit\'e F-67084 Strasbourg Cedex, France}
%\affiliation{$^d$Department of Atomic Physics, Lor\'and E\"otv\"os University,
%Budapest, Hungary}
\date{\today}

\begin{abstract} In this letter, we reconsider the Born-Infeld approach by replacing the Faraday's field-strength tensor by a complex one in the lagrangian density. We show that an extension of the least action principle to complex-valued lagrangian densities permits to understand why experiments have never exhibited nonlinear Born-Infeld effects.

\end{abstract}

%Use showkeys class option if keyword
\maketitle

%\section{Introduction}

In classical electrodynamics, two different theoretical approaches have dominated the comprehension of the
electromagnetic field (EMF) and of the electron. The first one resulting from the progress in the 19th century was proposing a clear distinction between the EMF and its sources, even if some topological problems, for instance the Dirac's monopole, have yielded to the necessity to the mutual quantization of the electric and magnetic charges.
The second one, justified and formalized later by the Quantum Field Theory (QFT), was rising the EMF as the only physical concept, letting the particles being only its singularities.\\
In the first approach, one of the remaining obstacles was the contradiction between the electron mass finiteness and the divergence of the electrostatic energy it has created.
Born and Infeld have proposed a new model, the so-called Born-Infeld Theory (BIT) in order to regularize the electrostatic energy without giving up the ponctual electron concept. This model yields to a deep modification of EMF motion equations (Maxwell's ones) making them non-linear and giving a free EMF. 
The BIT defines an arbitrary parameter permitting at the short spatial limit to recover the Maxwell's Theory and its linearity.
However, and even if its capability to integrate the Unified Theories was admitted, the BIT was suffering not to be supported by experimental results giving proofs for the non-linear effects predicted by the BIT.
The advent of quantum electrodynamics and of the Renormalization Theory have made the BIT without object while Quantum Field Theory was in progress.\\
In the framework of linear electrodynamics, one uses usually the Faraday tensor $(\Fij ij)$ and its Hodge-dual $(\dFij ij)$ to describe the electromagnetic field $(\mathbf{E},\mathbf{B})$. The  Maxwell's equations can be obtained from both tensors by mean of the lagrangian density $\mathcal{L}$
\beqn
\label{L_real}
\mathcal{L} =-\frac{1}{4}\mathcal{F}_{\mu\nu}\mathcal{F}^{\mu\nu}-j_\mu A^\mu= \frac{1}{4}\mathcal{F^*}_{\mu\nu}\mathcal{F^*}^{\mu\nu}-j_\mu A^\mu
\eeqn
which uses the Lorentz's invariant $\mathbf{E}^2-\mathbf{B}^2$. Motivated for building a nonlinear electrodynamics theory, Born and Infeld were looking for a covariant action by using the space-time metric tensor $g$ and the Faraday's tensor $\mathcal{F}$. The most general lagrangian is therefore  linear combination of $\sqrt{-\det g}$ and $\sqrt{-\det g + {\mathcal{F}\over k}}$ since $\sqrt{-\det \mathcal{F}}$ is not contributing to the motion equations. If one requires that for $\vert \mathbf{E} \vert$ and  $\vert \mathbf{B} \vert$ very smaller than $k$, one retrieves the quadratic lagrangian \eq{L_real}. This result is unique and corresponds to the lagrangian density developped in \cite{Born_Infeld_1933,Born_Infeld_1934,Born_1937}
\beqn
\label{L_BI}
\mathcal{L}_{BI}= -\frac{k^2}{2}
                              \bigg(\sqrt{-\det(g^{\mu\nu}+\frac{1}{k}\mathcal{F}^{\mu\nu})}-\sqrt{-\det(g^{\mu\nu})}\bigg)
\eeqn
 where $g_{\mu\nu}=\trm{diag}(1,-1,-1,-1)$ is the metric tensor, and $k$ is dimensionnal parameter. It uses the second Lorentz's invariant, $\mathbf{E}\cdot\mathbf{B}$, as it can be shown on the following equivalent expression

\beqn
 \mathcal{L}_{BI}=-k^2
                      \bigg(\sqrt{1-\frac{\bE^2-\bB^2}{k^2}
                                   -\frac{(\mathbf{E}\cdot\mathbf{B})^2}{k^4}}-1\bigg)
\eeqn
 This density describes a non-interacting gauge theory but has not been validated by experiments in order to exhibit non-linear classical effects  \cite{Rafelski}.  However, its introduction by Dirac to describe the electron and the muon by mean of a charged and closed membrane \cite{Dirac_1962,Leigh_1989}, the so-called Dirac-Born-Infeld became essential for the description of membranes and superstrings \cite{CM_1998,G_1998,Moreno_1998, Thor_1998,Yang_2000}.\\
In this letter, we reconsider the Born-Infeld approach by starting from the assumption that the right field-strength tensor which is used in \eq{L_real} and \eq{L_BI} has to be the complex Faraday tensor $\mathcal{F_C}=\mathcal{F}+i\cdot\mathcal{F^*}$, corresponding to the Lorentz's invariant $(\mathbf{E}+i\mathbf{B})^2$  \cite{Landau}. Those complex tensor $\mathcal{F_C}$ and this complex vector $\mathbf{F}=\mathbf{E}+i\mathbf{B}$ have already a long history since it has been introduced in 1907s' by L. Silberstein  \cite{Silberstein_1907,Silberstein_1924}. In particular, E. Schroedinger \cite{Schroedinger_1935} and P. Weiss \cite{Weiss_1936} have already proposed the use of this complex vector $\mathbf{F}$ in the framework of the Born-Infeld Theory, the so-called "New Field Theory".\\
By replacing $\mathcal{F}$ by $\mathcal{F_C}$ in \eq{L_real} and \eq{L_BI}, one defines respectively complex-valued lagrangian densities $\mathcal{L}_C$ and $\mathcal{L}_{BIC}$.

Since the extremum of a complex-valued functionnal was not well-defined, it was therefore not possible to exploit efficiently those complex-valued  lagrangian densities. Some recent works have filled this lack by giving a well-suited mathematical definition of a complex-valued  functionnal extremum  \cite{Gondran1,Gondran2,Gondran_Hoblos_2003}. \\
Later we shall use those results in order to define an extension of the least action principle to complex-valued lagrangian densities. Afterward, we show how from that definition, one can deduce Maxwell's equations with the complex Faraday tensor $\mathcal{F_C}$ and covariant complex lagrangian density. In the following paragraph, we still retrieve Maxwell's equations if one uses the complex Faraday's tensor in the Born-Infeld lagrangian density rather than the real one.
Finally, we conclude that the complex Faraday tensor $\mathcal{F_C}$ seems to be the relevant field-strength  tensor.

%\section{Analytic Lagrangian density for the Maxwell's equations}

The complex variationnal calculus is build on the definition of the complex-valued functions minimum~ \cite{Gondran1,Gondran2}. For a function $f$ from $\mathbb{C}^{n}$ to $\mathbb{C}$ which can be written as $f({\bf z}) \equiv f({\bf x},{\bf y})=P( {\bf x},{\bf y}) +iQ({\bf x},{\bf y}) $, one defines its minimum  with ${\bf z}_{0}={\bf x}_{0}+i{\bf y}_{0}$ such as
\beqn
\lbl{minmax}
\min\limits_{{\bf x}}\max\limits_{{\bf y}} P( {\bf x},{\bf y})=P({\bf z}_0) =\max \limits_{{\bf y}}\min\limits_{{\bf x}} P( {\bf x},{\bf y}).
\eeqn
This means that $({\bf x}_0,{\bf y}_0)$ is a saddle-point of $P({\bf x},{\bf y})$. A necessary condition for $z_0$ to be a minimum of an analytic function $f$ is $f^{\prime }({\bf  z}_{0}) =0$. A relevant example for our topic is the function from $\mathbb{C}$ to $\mathbb{C}$, $f(z)\equiv z^2=x^2-y^2+2ixy$ which is analytic. Its minimum is such as
\be
z_0^2=\min\limits_{x}\max\limits_{y}\{x^2-y^2\}=\min\limits_{x}\{x^2\}=0.
\ee
which yields to the right solution $z_0=0$. This example is important because the two quantities $x^2-y^2$ and $xy$ which are appearing respectively in the real and the imaginary part of $z^2$ are analogous to $\bE^2-\bB^2$ and $\mathbf{E}\cdot\mathbf{B}$.\\
One can define a variationnal calculus for complex-valued functionnals by extending those definitions. Thus, the Euler-Lagrange equation for a complex lagrangian density is the same as for the real one.

In the framework of classical linear electrodynamics, those equations are simply the  Maxwell's equations. One has to find out the extremum of the action  $\int\mathcal{L}d^4x$ according to the quadrivector $(\varphi, \mathbf{A})$
\beqn\label{nn}
    \mathcal{L}= \frac{1}{2}(\mathbf{E}^2
                   -\mathbf{B}^2)
                   -\rho\varphi+{\mathbf{j}}\mathbf{A}
\eeqn
with $\mathbf{E}$ and $\mathbf{B}$ relied to  $(\varphi,\mathbf{A})$ by
\beqn\label{E-et-B}
    \mathbf{E}= -\mathbf{\nabla}\varphi
                        -\frac{\partial\mathbf{A}}{\partial t}
    \ \ \trm{and}\ \
    \mathbf{B}= \mathbf{\nabla}\times\mathbf{A}
\eeqn
which yields to
\beqn\label{Maxwell-1ieme-Groupe}
    \mathbf{\nabla}\cdot\mathbf{B}=0
    \ \ \trm{and}\ \
    \mathbf{\nabla}\times\mathbf{E}
    +\frac{\partial\mathbf{B}}{\partial t}=0
\eeqn
Let's make the assumption that an alternative way to describe the fields which is more convenient  and elegant is to consider either the complex vector $\mathbf{F}=\mathbf{E}+i\mathbf{B}$ or the tensor $\mathcal{F_C}=\mathcal{F}+i\cdot\mathcal{F^*}$. This defines a complex-valued lagrangian density $\mathcal{L}_C$,
\beqn
\label{action-complexe}
    \mathcal{L}_C=-\frac{1}{4}\mathcal{F_C}_{\mu\nu}\mathcal{F_C}^{\mu\nu}-j_\mu A^\mu=\hf\mathbf{F}^2-\rho\varphi+{\mathbf{j}}\cdot\mathbf{A}.
\eeqn
It is possible to generalize the classical least action principle to this complex-valued field and to exhibit  an Euler-Lagrange-like equation by mean of complex variationnal calculus that we have defined in \eq{minmax} in order to find out the minimum of the complex action $\int\mathcal{L}d^4x$ for $\mathbf{F}$ defined with $(\varphi,\mathbf{A})$ as follow~:
\beqn\label{F-et-phi-A}
    \mathbf{F}=-\frac{\partial\mathbf{A}}{\partial t}
               -\mathbf{\nabla}\varphi
               +i\mathbf{\nabla}\times\mathbf{A}.
\eeqn

Since $\mathcal{L}$ is the real part of the complex lagrangian density $ \mathcal{L}_C=\mathcal{L}+iQ$, according to \eq{minmax}, the generalized complex least action principle applied to $\mathcal{L}_C$ yields to the same solution as  the classical least action principle applied to $\mathcal{L}$. This gives an equivalent form of the Maxwell's equations $\mathbf{F}$~:
\beqn\label{Maxwell-F}
    \mathbf{\nabla}\cdot\mathbf{F}=\rho
    \ \ \ \ \ et \ \ \ \
    \mathbf{\nabla}\times\mathbf{F}
    -{i}\frac{\partial\mathbf{F}}{\partial t}={i}\mathbf{j}.
\eeqn

One remarks that the computation of the complex Faraday tensor determinant gives $\det(\mathcal{F_C})=-\mathbf{F}^4$. This yields to $\sqrt{-\det(\mathcal{F_C})}=\mathbf{F}^2$. Thus, $\mathcal{L}_C$ can be therefore written as

\beqn
\label{complex_action}
    \mathcal{L}_C=\hf \sqrt{-\det(\mathcal{F_C})}-j_\mu A^\mu
    \eeqn

This shows that the complex Faraday's tensor seems to be the right field-strength tensor because it permits to define the right electrodynamics lagrangian by using the general covariance of the action.
If one tries to put in \eq{complex_action} the Faraday tensor (or its dual) rather than the complex one, one will not find the right lagrangian density because $\det(\mathcal{F})=\det(\mathcal{F^*})=(\mathbf{E}\cdot\mathbf{B})^2$.
%\section{Born-Infeld Theory}
The simultaneous mean ot the complex Faraday's tensor and of the covariance principle permits us to revisit the non-linear Born-Infeld Theory. We have seen in the previous section that if we consider that the relevant field-strength tensor in electrodynamics is the complex Faraday's one, the Born-Infeld lagrangian density should be rewritten by replacing $\mathcal{F}$ in \eq{L_BI} by $\mathcal{F_C}$.

\beqn
\lbl{L_BIC}
\mathcal{L}_{BIC}= -\frac{k^2}{2}
                              \bigg(\sqrt{-\det(g^{\mu\nu}+\frac{1}{k}\mathcal{F_C}^{\mu\nu})}-\sqrt{-\det(g^{\mu\nu})}\bigg)\nn
\eeqn

A straight calculation leads to
\begin{eqnarray*}
    \det\bigg(g^{\mu\nu}+\frac{1}{k}\mathcal{F_C}^{\mu\nu}\bigg) & = &\det\left( \begin{array}{cccc}
      1             & -\frac{F_x}{k}  & -\frac{F_y}{k}  & -\frac{F_z}{k}  \\
      \frac{F_x}{k} & -1              & i\frac{F_z}{k}  & -i\frac{F_y}{k} \\
      \frac{F_y}{k} & -i\frac{F_z}{k} & -1              & i\frac{F_x}{k}  \\
      \frac{F_z}{k} & i\frac{F_y}{k}  & -i\frac{F_x}{k} & -1              \\

    \end{array}
    \right)
    \nn
    & =& - (1-\frac{1}{k^2}\mathbf{F}^2)^2\nn
\end{eqnarray*}
We obtain therefore

\[
    \mathcal{L}_{BIC}=-\frac{k^2}{2}[(1-\frac{1}{k^2}\mathbf{F}^2)-1]
                       =\frac{1}{2}\mathbf{F}^2.
\]

This expression shows first that the Born-Infeld complex lagrangian is the complex Faraday's one. Second, it shows that the Euler-Lagrange equations obtained from it, will give no linear effects. It is in agreement with experiments  \cite{Rafelski}. This means that there are no nonlinear effects excepted in the quantum treatment of electrodynamics.The difference between the Born-Infeld complex approach and the one presented in this letter is due to the fact that the real part of $\mathcal{L}_{BIC}$ is different from the $\mathcal{L}_{BI}$ one. Indeed, the real part of the square root of a complex number is not equal to the square root of its real part.

%\section{Conclusion}
Construction of a Field Theory needs to choose the right tensorial field for the right lagrangian density. In electrodynamics, Born and Infeld have choosen respectively the Faraday's field-strength tensor $(\Fij ij)$ and the lagrangian density $\mathcal{L}_{BI}$.
Experiments did not agree with nonlinear effects predicted by their theory. They have stated in 1934  \cite{Born_Infeld_1934} that \emph{"which of these action principles is the right one can only be decided by their consequences"}. Following that statement, we have showed that if they have defined a good candidate to be a lagrangian density, the right field-strength tensor is the complex Faraday's one in that case. Using an appropriate extension of the least action principle on both $\mathcal{L}_C$ and $\mathcal{L}_{BIC}$ and using the tensor $\mathcal{F_C}$, one retrieves the  Maxwell's equations.\\
The quantity $\mathbf{F}=\mathbf{E}+i\mathbf{B}$ has been viewed often as a possible photon wave function as in the interesting review  \cite{Birula_1996}. It is a pragmatic way to consider the existence of this wave function, as stated by P.A.M. Dirac in 1958  \cite{Dirac_1958} : \emph{"The essential point is the association of each of the translational states of the photon with one the wave functions of ordinary wave optics"}.\\ 
The complex lagrangian density proposed in this letter is therefore an explicit functionnal of the wave function. It gives a roadmap to a possible generalization and link between general relativity and electrodynamics. The challenge will be to define a covariant action by combining the metric tensor and a generalized complex Faraday tensor to a curved space.\\
We thank Professor Janos Polonyi for useful and helpful discussions.

%\section{Aknowledgements}

\end{document}